\documentclass[12pt]{article}
\usepackage{amsmath,bm}
\usepackage{amssymb}
\usepackage{graphicx}
\usepackage{amsfonts}         
\usepackage{fancybox}   
\usepackage{yfonts}

\renewcommand{\baselinestretch}{1.4}
\def\comments#1{}

\def\p{\partial}

\def\Im{{\rm Im\hskip0.1em}}

\def\vev#1{\langle{#1}\rangle}

\def\CA{{\cal A}}

\def\CE{{\cal E}}

\def\CT{{\cal T}}

\def\CN{{\cal N}}
\def\CO{{\cal O}}
\def\CP{{\cal P}}

\def\a{\alpha}

\def\II{\relax{I\kern-.10em I}}

\def\IB{\relax{\rm I\kern-.18em B}}
\def\ID{\relax{\rm I\kern-.18em D}}
\def\IE{\relax{\rm I\kern-.18em E}}
\def\IF{\relax{\rm I\kern-.18em F}}
\def\IG{\relax\hbox{$\inbar\kern-.3em{\rm G}$}}
\def\IGa{\relax\hbox{${\rm I}\kern-.18em\Gamma$}}
\def\II{\relax{\rm I\kern-.18em I}}
\def\IK{\relax{\rm I\kern-.18em K}}

\def\inbar{\,\vrule height1.5ex width.4pt depth0pt}

\def\p{\partial}

\def\frac#1#2{{#1 \over #2}}

\newdimen\tableauside\tableauside=1.0ex
\newdimen\tableaurule\tableaurule=0.4pt
\newdimen\tableaustep
\def\phantomhrule#1{\hbox{\vbox to0pt{\hrule height\tableaurule width#1\vss}}}
\def\phantomvrule#1{\vbox{\hbox to0pt{\vrule width\tableaurule height#1\hss}}}
\def\sqr{\vbox{%
  \phantomhrule\tableaustep
  \hbox{\phantomvrule\tableaustep\kern\tableaustep\phantomvrule\tableaustep}%
  \hbox{\vbox{\phantomhrule\tableauside}\kern-\tableaurule}}}
\def\squares#1{\hbox{\count0=#1\noindent\loop\sqr
  \advance\count0 by-1 \ifnum\count0>0\repeat}}
\def\tableau#1{\vcenter{\offinterlineskip
  \tableaustep=\tableauside\advance\tableaustep by-\tableaurule
  \kern\normallineskip\hbox
    {\kern\normallineskip\vbox
      {\gettableau#1 0 }%
     \kern\normallineskip\kern\tableaurule}%
  \kern\normallineskip\kern\tableaurule}}
\def\gettableau#1 {\ifnum#1=0\let\next=\null\else
  \squares{#1}\let\next=\gettableau\fi\next}

 \def\eqnn#1{\xdef #1{(\secsym\the\meqno)}\writedef{#1\leftbracket#1}%
 \global\advance\meqno by1\wrlabeL#1}
 \def\eqna#1{\xdef #1##1{\hbox{$(\secsym\the\meqno##1)$}}
 \writedef{#1\numbersign1\leftbracket#1{\numbersign1}}%
 \global\advance\meqno by1\wrlabeL{#1$\{\}$}}
 \def\eqn#1#2{\xdef #1{(\secsym\the\meqno)}\writedef{#1\leftbracket#1}%
 \global\advance\meqno by1$$#2\eqno#1\eqlabeL#1$$}

\global\newcount\itemno \global\itemno=0

\def\itemaut#1{\global\advance\itemno by1\noindent\item{\the\itemno.}#1}

\def\del{\partial}

\def\({\left(}
\def\){\right)}

\def\eg{{\it e.g.}}
\def\ie{{\it i.e.}}

\newif{\ifeq}           
\eqtrue                 
\let\non\nonumber

\newcommand{\be}{\begin{equation}}
\newcommand{\ee}{\end{equation}}
\newcommand{\bea}{\begin{eqnarray}}
\newcommand{\eea}{\end{eqnarray}}
\newcommand{\bean}{\begin{eqnarray*}}
\newcommand{\eean}{\end{eqnarray*}}

\newcommand{\bb}{\mathbb}

\def\({\left(}
\def\){\right)}
\def\[{\left[}
\def\]{\right]}

\newcommand{\C}[1]{$(\ref{#1})$}

\newcommand{\half}{\frac{1}{2}}

\newcommand{\rt}{{\sqrt 2}}

\renewcommand{\a}{\alpha}
\renewcommand{\b}{\beta}

\newcommand{\g}{\gamma}

\renewcommand{\r}{\rho}
\newcommand{\s}{\sigma}
\renewcommand{\l}{\lambda}

\newcommand{\A}{{\cal A}}

\newcommand{\N}{{\cal N}}
\renewcommand{\O}{{\cal O}}

\newcommand{\K}{{\cal K}}

\def\CO{\O}

\newcommand{\IP}{{\mathbb P}}

\newcommand{\IZ}{{\mathbb Z}}

\def\ie{{\it i.e.}}

\newcommand{\lsim}{\,\raise.3ex\hbox{$<$\kern-.75em\lower1ex\hbox{$\sim$}}\,}
\newcommand{\gsim}{\,\raise.3ex\hbox{$>$\kern-.75em\lower1ex\hbox{$\sim$}}\,}

\def\p{\partial}

\newcommand{\scz}{\setcounter{equation}{0}}

\newif{\ifeq}
\eqtrue

\def\KK{(1-\b^{2}\r^{2}g)}

\def\KK{K}

\def\CK{{\cal K}}

\begin{document}

\begin{titlepage}

\begin{flushright}
MIT-CTP/3962
\end{flushright}
\vfil

\begin{center}
{\huge Hot Spacetimes for Cold Atoms}\\
\end{center}
\vfil
\begin{center}
{\large Allan Adams, Koushik Balasubramanian, and John McGreevy}\\
\vspace{1mm}
Center for Theoretical Physics, MIT,
Cambridge, Massachusetts 02139, USA\\
{\tt awa at mit.edu}\\
\vspace{3mm}
\end{center}

\vfil

\begin{center}

{\large Abstract}
\end{center}

\noindent
Building on our earlier work and that of Son,
we construct string theory duals of 
non-relativistic critical phenomena at finite temperature and density.
Concretely, we find black hole solutions of type IIB supergravity 
whose asymptotic geometries realize the Schr\"odinger group as 
isometries.
We then identify the non-relativistic conformal field theories to which they are dual.
We analyze the thermodynamics of these black holes,
which turn out to describe the system at finite 
temperature and finite density.
The strong-coupling result for the shear viscosity
of the dual non-relativistic field theory 
saturates the KSS bound.
%
\vfill
\begin{flushleft}
July 2008
\end{flushleft}
\vfil
\end{titlepage}
\newpage
\renewcommand{\baselinestretch}{1.1}  

\renewcommand{\arraystretch}{1.5}

\section{Introduction}\scz
%

The hydrodynamics of cold atoms with interactions at the unitarity limit 
(for reviews, see \cite{zwerger})
is a subject crying out for an effective strong-coupling description.
Experimentally, these systems are under extensive and detailed study, and rich data exist.
For example, the shear viscosity has been extracted \cite{Schafer:2007pr} from 
energy-loss during sloshing experiments \cite{THOMAS}, leading to an estimate 
for the ratio $\eta \over s$ which approaches
the bound conjectured by 
\cite{KovtunDE}.  While this bound is universal
in {\it relativistic} systems with classical gravity duals
\cite{buchelliu, KovtunDE}, 
the system of cold fermionic atoms at unitarity
is most certainly {\em not} a relativistic one,
though it does have non-relativistic conformal symmetry
(for a systematic discussion, see \cite{Nishida:2007pj}).
One could imagine that the 
nonrelativistic nature of the system
has an important effect on this ratio.
Indeed, the counterexamples to the $\eta \over s$ bound 
proposed in \cite{Cohen}
arise in nonrelativistic systems.
Theoretically, however, these systems are hard: perturbative techniques are inadequate, and lattice methods are 
difficult to apply to dynamical questions 
(though see \cite{harvey}).  

In the relativistic context, an effective strong-coupling description of a CFT can sometimes 
be found in terms of a gravitational theory in extra-dimensional spacetimes
\cite{MaldacenaRE, witten} in which the conformal symmetry of the CFT arise as the isometries of the geometry.  
Via the Hawking phenomenon, the thermal ensemble of such systems
is constructed by placing a black hole
in the extra-dimensional geometry \cite{witten}.
The rigid structure of
black hole spacetimes, when combined
with the finite-temperature
gauge/gravity duality,
has led to the observation of
universal behavior of these strongly coupled gauge theories
at finite temperature.
For a nice review of this work, see
\cite{Son:2007vk}.


In the non-relativistic case, a natural guess for a strong-coupling description is
a dual geometry whose isometries reproduce the symmetries of the non-relativistic CFT (NRCFT).
Such geometries were constructed recently in
\cite{Son:2008ye, Balasubramanian:2008dm}, with
the metric taking the form\footnote{For previous 
appearances of related spacetimes 
in the pre-gauge/gravity-duality literature,
see \cite{bateman,Duval,Horvathy}.  
For studies of the supersymmetrization of the Schr\"odinger group 
please see \cite{Sakaguchi:2008rx}.
See also \cite{Wen:2008hi}.
}
\be
\label{pappalardometric}
ds^2 = L^2 \( - {2 \beta^{2} dt^2 \over r^{2z} } + { 2 d\xi dt + d \vec x^2  + dr ^2 \over r^2} \).
\ee
Here, $\vec x$ is a vector of $d$ spatial coordinates and $z$ is the dynamical exponent, which takes the value $z=2$ for the
fermions at unitarity.
In \cite{Son:2008ye, Balasubramanian:2008dm}
this spacetime was shown to solve the equations 
of motion of Einstein gravity coupled
to a gauge field of mass $m_A^2 = {z(z+d)\over L^2}$ and 
a cosmological constant $ \Lambda = { (d+1)(d+2) \over L^2 } $,
and was argued to be dual to an NRCFT at zero temperature and zero density.
To 
study via duality an
NRCFT at finite temperature and density, then,
we need to put a black hole inside this geometry.


In this paper we will construct black holes 
with the asymptotics of (\ref{pappalardometric}),
show that they arise as 
solutions of string theory,
and identify the specific non-relativistic conformal field theories
to which they are dual.
An analysis of the thermodynamics of these black hole spacetimes
shows that they describe the dual non-relativistic CFTs
at finite temperature and finite density, with the previously-studied 
geometry of \cite{Balasubramanian:2008dm,Son:2008ye}
arising in the zero temperature, zero density limit.
Along the way we identify a scaling limit which describes
the system at zero temperature but nonzero density.
To produce these solutions, we will use a solution-generating technique
called the 
Null Melvin Twist \cite{Bergman:2001rw, Alishahiha:2003ru, Gimon:2003xk,
Hubeny:2005qu, Hubeny:2005pz},
which we will review in detail below.
This Melvinizing procedure is the sought-for
analog of the plane wave limit
described in 
the introduction and conclusion of 
\cite{Balasubramanian:2008dm}.


We should emphasize at this point that the
NRCFT describing Lithium atoms tuned to a Feschbach resonance
probably does not literally have a weakly-coupled gravity dual.
However, the Lithium system has closely related cousins
which do have 't Hooft limits -- indeed,  
the NRCFTs dual to our black hole spacetimes are
precisely such creatures.
Our hope is that these ideas will
be useful for studying strongly-coupled cold atoms
in at least the same sense
in which the $\CN=4$ theory
has been useful for studying universal properties
of strongly-coupled relativistic liquids,
including those made out of QCD.

The paper is organized as follows.
After reviewing
the correspondence 
proposed in \cite{Son:2008ye, Balasubramanian:2008dm} 
in section 2,
we show in section 3 that it can be embedded in string theory.
The solutions are constructed using
the Null Melvin Twist, 
a machine which eats supergravity solutions
and produces new ones.  The machine has several dials, which we will tune to various ends.
The input solution that produces the metric (\ref{pappalardometric}) 
is the extremal D3-brane in type IIB. 
When (in section 3.3) we feed to the Melvinizing machine
the near-extremal D3-brane,
we find that it produces black brane solutions
which asymptote to the spacetimes (\ref{pappalardometric}).
We then provide some rudimentary understanding of the identity of 
the theory at weak coupling in this realization.
In section four we analyze the thermodynamics.
In section five we compute the shear viscosity,
and show that the strong-coupling universality found by \cite{buchelliu, KovtunDE}
extends beyond the class of relativistic systems.
We conclude with a discussion of interesting open questions.
The appendices contain the details of the Melvinization
process, an argument for frame-independence of the viscosity calculation,
and some progress towards a 5d effective action.

\section{Schr\"odinger Spacetimes and Non-relativistic CFTs} \scz

Non-relativistic systems which enjoy conformal invariance in $d$ spatial dimension are governed by a symmetry algebra known as the d-dimensional Schr\"odinger algebra.  In addition to the usual generators of the Galilei group, \ie\ generators of spatial translations, $P_{i}$, rotations $M_{ij}$, Galilean boosts $K_{i}$, and time translations, $H$, the Schr\"odinger algebra includes a dilatation operator, $D$, and a number operator, $N$, whose non-trivial commutators are
$$
[D,P_{i}]=iP_{i} ~~~
[D,K_{i}]=i(1-z)K_{i} ~~~
[D,H]=izH ~~~
[D,N] = i ( 2-z) N ~~~
[P_{i},K_{j}]=-\delta_{ij}N
$$
where $z$, the ``dynamical exponent'', determines the relative scaling between the time-coordinate and the spatial coordinates, $[t]$=${\rm length}^{z}$.  In the special case $z=2$, the algebra may be extended by an additional ``special conformal'' generator, $C$, whose non-trivial commutation relations are
$$
[D,C]=-2iC ~~~ [H,C]=-iD.
$$
In this case $z=2$,
both $D$ and $N$ may be diagonalized, so representations of the Schr\"odinger algebra are in general labeled by two numbers, a dimension $\Delta$ and a ``number'' $\ell$.  
For
fermions at unitarity, this number is precisely the fermion number.

Motivated by the relativistic AdS/CFT correspondence, it is natural to wonder whether there exists gravitational duals for non-relativistic CFTs.  By analogy to the relativitstic case, we expect such a gravitaional description to realize the symmetry group of the CFT as the isometry group of the spacetime.  However, since there are now {\em two} symmetry generators which may be diagonalized and whose eigenvalues label inequivalent representations (in the AdS case, there is only one, the dimension), we may expect any spacetime which has the Schr\"odinger algebra linearly realized as its isometry group to be two dimensions higher than the CFT, as opposed to one-dimension higher as in the case of AdS.

Such geometries were explicitly constructed in \cite{Balasubramanian:2008dm} and \cite{Son:2008ye}.  More precisely, these papers constructed a $d+3$-dimensional metric realizing the $d$-spatial-dimensional Schr\"odinger group as its isometry group, and conjectured the associated gravitational system to be dual to non-relativistic CFTs at zero temperature and zero density.  The metric they presented appears in \C{pappalardometric} above.  We will refer to these metrics as $Sch^{z}_{d+3}$, where $z$ labels the dynamical exponent and $d$ the number of spatial dimensions (note that d+3 = (d+1)+2); in the special case $z=1$, $Sch^{1}_{d+3}=AdS_{d+3}$.  
Many of the generators are simple to realize as isometries of this geometry.  For example, the dilatation $D$ is realized as the simultaneous scaling,
$$
\{t,\xi,\vec{x},r\} ~~ \stackrel{\l D}{\to} ~~ \{\l^{z}t,\l^{2-z}\xi,\l\vec{x},\l r\},
$$
while a boost $K$ acts as,
$$
\vec{x}\to \vec{x}-\vec{v}t ~~~~~~ \xi \to \xi +\vec{v}\cdot\vec{x}-{v^{2}\over 2}t.
$$
Considerably less obvious, but extremely important to what follows, is the identification,
$$
N = i\p_{\xi}.
$$
The fact that the number operator in a non-relativistic conformal field theory is gapped (one Li atom, two Li atoms, three...) tells us that $\xi$ must be periodic.  But $\xi$ is a null direction in the bulk geometry.  As such, we appear to be forced into a 
discrete light cone quantization (DLCQ).  This will be made more precise in Section 3.

At first sight, compactifying $\xi$ may look problematic.  
For example, this may appear to violate boost invariance.
However, boost invariance remains unbroken precisely because the $\xi$ direction is null;
this follows from the commutator $[\hat N, \hat K_i ] = 0$ in the Schr\"odinger algebra\footnote{We thank Simon Ross for clarifying this issue.}.
Perhaps more troublingly, compactifying $\xi$ would appear to introduce a null conical singularity at $r\to\infty$, which suggests that our metric should not be reliable in the strict IR.
However, this singularity is unphysical.  As we shall see below, the singularity goes away as soon as we turn on any finite temperature -- the would-be null singularity is lost behind a finite horizon which shrouds a garden-variety schwarzschild singularity.  Meanwhile, physically, we always have some finite $T$ in a realistic cold-atom system, and thus a natural IR regulator.  Finally, and most sharply, even in the strict $T\to0$ case, the dynamics will resolve this ``singularity'' in a fashion familiar from the study of null orbifolds of flat space \cite{Lawrence:2002aj,Liu:2002kb,Fabinger:2002kr,Horowitz:2002mw}: a pulse of stress-energy sent towards large $r$ is steadily blue-shifted until its back-reaction is no longer negligible; analysis of the back-reaction then shows that the would-be null-singularity turns over into a spacelike singularity shrouded behind a (microscopic) horizon.  All of which is to say, the strict $T\to0$ limit of our NRCFT is unstable to thermalization upon the introduction of any energy, no matter how small.  The challenge, in both the NRCFT and the dual spacetime, is not to crank up a finite temperature, but to drive the temperature low.

Importantly, this metric is not a vacuum solution of the Einstein equations.  As a result, it is necessary to 
couple the system to additional background fieldstrengths (a pressureless dust and a negative cosmological constant) whose stress tensors cancel the non-zero Einstein tensor of the spacetime metric\footnote{The recent papers \cite{Goldberger:2008vg, Barbon:2008bg}\ find a solution of the vacuum einstein equations with only a cosmological constant -- this is just the DLCQ of AdS, with the periodic identification breaking the AdS symmetry group to its 
Schr\"odinger subgroup.  As we will discuss in considerably more detail in the next section, this corresponds to a degenerate limit of the backgrounds considered in \cite{Balasubramanian:2008dm,Son:2008ye} and, more generally, in this paper.}.  As we shall see in the next section, this system -- a metric with Schr\"odinger isometries supported by background fieldstrengths for massive tensor fields -- has a natural embedding into string theory.

Further evidence for the conjectured duality is provided by a comparison of Green functions for scalar operators
as computed in the NRCFT and gravity.  The $n$-point Green's function of an operator ${\cal O}$ in a NRCFT, is
determined by the scaling dimension $\Delta_{\cal O}$ and the particle number
$N_{\cal O}$ \cite{Nishida:2007pj}. 
Indeed, as shown by Nishida and Son \cite{Nishida:2007pj},
this is the case in any
nonrelativistic CFT, since that's what's required to
specify a representation of the $z=2$ Schr\"odinger algebra.
The two-point Green's function calculated using
the gravity theory \cite {Balasubramanian:2008dm} has the same 
form as that of 
\cite {Nishida:2007pj}.
Perhaps not surprisingly, the spectrum of the number operator
in the theories dual to geometries of the form (\ref{pappalardometric}) 
is the set of integers, since it arises from the tower
of Kaluza-Klein momenta in the $\xi$ direction.
The name for the $\xi$-momentum conservation law in the $n$-point functions
$$l_1 + l_2 + ... = l_1' + l_2 ' + ... $$
is ``Bargmann's superselection rule on the mass''
\cite{hagen}.
The stress tensor is an operator which commutes with
the particle number operator (this is a consequence of the 
Schr\"odinger algebra).  The fluctuations of the bulk metric dual to the 
stress tensor therefore have
zero $\xi$-momentum.

Time reversal is an antiunitary operation, which means that
it complex conjugates the wavefunctions.
In particular, say in a weakly coupled theory with
field operator $\Psi$, it acts on an operator $\CO_l$ by
$$
T: \CO_l = \Psi^l ... \Psi^{\dagger~l-k}
\mapsto \CO_{-l}. $$
This is consistent with the fact that our geometries
are preserved only by the combined operation
$$  t \mapsto - t,~~~  \xi \mapsto - \xi. $$

\section{Embedding in String Theory}\scz

In this section we will show that the zero-temperature solutions described above have a natural embedding into solutions of Type II string theory.  For simplicity, we will mostly focus on the case $z=2$, $d=2$, though many other cases 
may be equally straightforward.  These solutions may be generated in a number of equivalent ways.  One useful technique is the ``Null Melvin Twist,'' (which was named in \cite{Gimon:2003xk}) which will be described in detail momentarily.  We will use this technique to construct solutions which embed the $Sch^{2}_{5}$ geometry discussed above into string theory, as well as solutions which describe the system at finite temperature and chemical potential.  A second useful technique, which is in fact completely equivalent for the backgrounds we consider, is a simple modification of 
Discrete Light Cone Quantization; this presentation will make the structure of the dual field theory transparent.  Let's begin with the twist.

\subsection{The Null Melvin Twist}

The Null Melvin Twist is a solution generating technique for IIB supergravity which eats known solutions and spits out new solutions with inequivalent asymptotics.
Melvinization has largely been used to construct gravity duals of non-commutatitve and non-local field theories \cite{Bergman:2001rw, Alishahiha:2003ru, Gimon:2003xk, Hubeny:2005qu, Hubeny:2005pz}; in the case at hand, the Null twisting produces an extremely mild form of non-locality, that of non-relativistic theories with instantaneous interactions.  Importantly, these backgrounds have the property that all curvature scalars are identical to those of the original solution \cite{Alishahiha:2003ru}; as a result, the constraints on when the supergravity is reliable (\eg\ $\lambda\ll 1$) carry over directly.  This will also be clear from our analysis of the boundary field theory below.

Interestingly, some solutions with $Sch^{z}_{5}$ asymptotics have already appeared in the Melvinizing literature.  For example, the $T=0$ limit of one of the solutions in \cite{Alishahiha:2003ru, Hubeny:2005qu} corresponds to $Sch^{z=3}_{5}$, though the form field backgrounds break part of the symmetry group\footnote{We thank Mukund Rangamani for pointing out the previous appearance of these spacetimes in string theory.}.  These were described as dual to ``dipole theories'' with a non-trivial star-algebra in the dual field theories.  Here we will argue that these and some other backgrounds generated by the Null Melvin twist are in fact dual to NRCFTs.

The procedure itself is baroque but elementary.  The first step is to choose a IIB background with two marked isometries, which we will call $dy$ and $d\phi$.  We then (1) boost along $dy$ with boost parameter $\g$, which generates (in general) a new $dydt$ term in the metric,
(2) T-dualize a la Buscher\footnote{The full Buscher rules, and our conventions for them, are given in an appendix.} along $dy$, which generates a new $dy\wedge dt$ term in the NS-NS $B$-field and a non-trivial dilaton profile,
$g_{yy} \to  {1\over g_{yy}}$, $B_{t y} \to  {g_{t y}\over g_{yy}}$ and $\Phi \to \Phi -\half \ln{g_{yy}}$, 
(3) re-diagonalize our isometry generators by shifting $d\phi \to d\phi+\a dy$, which generates a new term in the metric of the form $ds^{2} \to ... +(d\phi+\a dy)^{2}$, 
then return to our original frame by 
(4) T-dualizing back along $dy$, which generates $dydt$ terms in the metric and $d\phi\wedge dy$ terms in $B$, and (5) boosting back along $dy$ with boost parameter $-\g$.   

All of this leaves us back in the original frame with a new metric, $B$-field and non-trivial dilaton, all of which are horrendously complicated functions of the two knobs, $\g$ and $\a$.  The final step of the Null twist is to (6) simplify this morass by taking a scaling limit in which the boost becomes infinite, $\g\to\infty$, and the twisting goes to zero, $\a\to0$, such that the product $\half \a e^{\g}=\b$ remains finite.  The result is a new solution of the full IIB equations of motion with non-trivial background NS-NS 2-form and deformed metric with asymptotics inequivalent to the original solution.

\def\dz{{(d\chi +\A)}}

\subsection{Rampaging Melvin Eats Extremal D3-brane, Spits Out $Sch^{2}_{5}$}

Let's apply this procedure to our canonical example, the extremal D3-brane, a solution of IIB supergravity with metric,
$$
ds^{2} = {1\over h} \( -d\tau^{2} +d\vec{x}^{2} \) + h \(d\r^{2} + \r^{2} ds_{S^{5}}^{2}  \)
$$
where $h^{2}= 1+{R^{4}_{A} \over \r^{4}}$ is the usual D3 harmonic function, and self-dual five-form flux
$$ 
F^{(5)} = {1\over r^5} d\tau \wedge dy \wedge dx_1 \wedge dx_2 \wedge dr 
+ \Omega_5 ~\! d\theta\wedge d\phi \wedge d\psi \wedge d\mu\wedge d\chi ,
$$
with $ \Omega_5 ={1\over 8} \cos \theta \cos\mu \sin ^3 \mu $. 
We must first choose the two isometry directions, $dy$ and $d\phi$, along which to Melvinize.  Let's take $dy$ to lie along the worldvolume and $d\phi$ along the $S^{5}$; without loss of generality, we can choose coordinates such that $y=x_{3}$.  A particularly convenient (though by no means the only possible) choice for $d\phi$ is given by the Hopf fibration $S^{1}\to S^{5}\to \IP^{2}$ with metric,
$$
ds^2_{S^5} = ds^2_{\IP^2} + (d\chi +\A)^2
$$
where $\chi$ is the local coordinate on the Hopf fibre and $\A$ is the 1-form potential\footnote{We can compute $\A$ from the K\"ahler potential  on $\IP^{2}$, $K = t \ln{\sum |z_i|^2}$, ie
$$
\A_{i} = ir{\bar{z}^{\bar{i}}\over\sum|z^{i}|^{2}}.
$$
For completeness, and because we had a pointlessly slow search for this data in the literature, we present an explicit set of conventions and coordinate systems in an appendix.} for the kahler form on $\IP^2$, ie $J_{\IP^{2}}=d\A$.  We thus take $d\phi=d\chi$.  Note that both $dy$ and $d\phi$ act freely, which is important for our solution to remain non-singular.  

Melvin being a very messy eater, we will hide the details of the procedure in the Appendix and simply write down the result, which is:
\bea
ds^2 &=& {1\over h}\(  -d\tau^{2}(1+\b^{2}\r^{2})  +dy^{2}(1-\b^{2}\r^{2})  +2d\tau dy(\b^{2}\r^{2})  \) +h \r^{2}\dz^2   \non \\
B&=&2\b\r^{2}\dz\wedge( d\tau +dy)   \non \\
\Phi &=& \Phi_{0} \non
\eea
Note that nothing has happend to the five-form along the way, since T-duality takes $d\Omega^{5}$, the top form on the sphere,  to $dy\wedge d\Omega^{5}$, so that the twist $d\phi \to d\phi + \b dy$ of step (3) acts trivially.  We thus have in our final solution the same five-form flux as in the beginning,
$$
F_{5} = (1+*) \Omega_{5} d\theta\wedge d\phi \wedge d\psi \wedge d\mu\wedge d\chi
$$

To locate the $Sch^{2}_{5}$ hiding inside this solution, a few more steps are helpful.  Changing coordinates to $t=(y+\tau)/\sqrt{2}$ and $\xi=(y-\tau)/\sqrt{2}$, our background becomes,
\bea
ds^2 &=& {1\over h}\( \b^{2}\r^{2}~\! dt^{2} +2dtd\xi  \) +h \r^{2}\dz^{2}  \non \\
B&=&2\b\r^{2}\dz\wedge dt ~~~~~~ \Phi = \Phi_{0} \non
\eea
Adding back in all the terms we dropped in the first step then gives,
\bea
ds^2 &=& {1\over h}\(
  -\b^{2}\r^{2}~\! dt^{2} +2dtd\xi + d\vec{x}^{2} \) +h \(d\r^{2} + \r^{2}ds_{S^{5}}^{2} \)  \non \\
B&=&2\b\r^{2}\dz\wedge dt ~~~~~~ \Phi = \Phi_{0} \non
\eea
Finally, taking the near-horizon limit, $h\to R^{2}_{A}/ \r^{2}$ and switching to the global radial coordinate $r= R_{A}^{2}/\r$, in terms of which $h={r^{2}\over R^{2}_{A}}$, the solution becomes,
\bea
ds^2 &=& {R_{A}^{2}\over  r^{2}}\[  -{2\Delta^{2}\over r^{2}}dt^{2} + 2dtd\xi + d\vec{x}^{2} + dr^{2}  \] + R^{2}_{A} ds_{S^{5}}^{2}   \non \\
B&=&2\rt~\!\Delta {R^{2}_{A} \over r^{2}} \dz\wedge dt    ~~~~~~~~
\Phi = \Phi_{0} \non .
\eea
Upon compactifying on the $S^{5}$, we precisely recover the Schr\"odinger geometry with $d=2$ and $z=2$, our sought-after $Sch^{2}_{5}$.

Note, too, the appearance of the parameter $\beta$, which was implicitly set to $1/\sqrt{2}$ in the earlier results of \cite{Son:2008ye, Balasubramanian:2008dm},
by a choice of units.  The utility of this parameter is considerable in what follows.  For now, note that retaining it allows a very revealing limit, \ie\ $\b\to0$, in which the solution above reduces to the extremal D3-brane solution with which we began.  This suggests that there should be a more intrinsic description of our solution as a garden-variety deformation of $AdS$; we will explore this relation later in this section.

A note on dimensions.  As discussed in Section 2, ensuring the quantization of the
spectrum of the Schr\"odinger number operator $\hat N$ requires compactifying the direction $\xi$, something not implemented in the Null Melvin Twist described above; this introduces a new dimensionful parameter to the game, the length scale $L_{\xi}$.  Meanwhile, $\phi$ is an angular direction along a compact space and so $d\chi$ is dimensionless, which means $\a$, and thus $\b$, must have dimensions of $1/{\rm length}$.   Our solutions would thus appear to have two dimensionful parameters, 
$L_{\xi}$ and $\b$.  In the case above, however, the specific values of $L_{\xi}$ and $\b$ can be rescaled by rescaling the coordinates as
$$ t \to \beta t, ~~ \xi \to \beta^{-1} \xi, $$
leaving only the product $\b \over L_{\xi}$ invariant.  It is this ability to scale away\footnote{Scaling away the value of $\b$ will not be possible in the finite-temperature solutions described below.} $\b$ which allowed 
\cite{Son:2008ye, Balasubramanian:2008dm} to set $\b=1/\sqrt{2}$.   Our 
extremal solutions are thus parameterized by a single dimensionless parameter, 
$\b \over L_{\xi}$.  Holding $L_{\xi}$ fixed while scaling $\b$ to zero gives a particularly trivial background which respects the Schr\"odinger group as its isometries; we shall return to this example below.

\def\ceff{c_{\rm eff}}

Relatedly, in the above we have set $c$ to 1; it is easy to reintroduce $c$ by taking $dt\to c~\!dt$.  Interestingly, the way we are getting a non-relativistic limit is {\em not} by taking $c\to\infty$; rather, the asymptotic geometry has an effective $\ceff \sim {c\beta\over r^{2}}$ which goes to $\infty$ as we approach the boundary.  In addition to $L_{\xi}$ and $\b$, then, our non-relativistic theory thus contains a finite velocity, $c$.  It is natural to interpret this velocity as the speed of sound, $v_{s}$.  
We will test this interpretation in the future.  
For now, we suppress factors of $c$, which are easy to restore.

We have thus constructed a solution of Type IIB string theory of the form Sch$^{2}_{5}\times S^{5}$ which is dual, according to the results of \cite{Balasubramanian:2008dm,Son:2008ye}, to some theory which respects the Schr\"odinger symmetry algebra, \ie\ a non-relativistic conformal field theory with $d=2$ and $z=2$.  
We note that it is straightforward to repeat the analysis of this section for other IIB backgrounds with the requisite isometries, including in particular other-dimensional branes and other choices of isometries along which to Melvinize.
In the next section, 
we use the same techniques
to construct a dual description of such an NRCFT at finite temperate and chemical potential.


\def\bb{\delta}
\def\KK{K}
\def\[{\left[}
\def\]{\right]}

\subsection{Solutions with Finite Temperature and Chemical Potential}

As we saw above, feeding the Melvinizing machine an extremal D3-brane produces 
a solution of IIB string theory with spacetime geometry $Sch^{z=2}_{5}\times S^{5}$ whose dual field theory is a NRCFT at zero temperature.  Experience with AdS/CFT suggests that putting the NRCFT at finite temperature should correspond to 
the introduction of a Rindler horizon
in the bulk spacetime, with modes of the boundary theory thermalized by the Hawking radiation of the black hole.  So we need to figure out a way to embed a non-extremal black hole inside our Schr\"odinger spacetime.  It is natural to guess that feeding Melvin a black D3-brane, which shares the asymptotic $AdS_{5}\times S^{5}$ structure of the extremal D3-brane, should produce a black hole spacetime which is asymptotically $Sch^{z=2}_{5}\times S^{5}$.  As we shall see, this is correct, with one important modification which will become clear after Melvin does his thing.  

Thus motivated, let's Melvinize the black D3-brane solution of IIB string theory.  The starting solution is,
$$
ds^{2} = {1\over h} \( -d\tau^{2}f +dy^{2}+d\vec{x}^{2} \) + h \({d\r^{2}\over f} + \r^{2}\[ds_{\IP^{2}}^{2} + (d\phi +\A)^2 \] \)
$$
where $h= {R^{2}_{A} \over \r^{2}}$ is the near-horizon limit\footnote{It is easy to keep the full geometry; we skip directly to the near horizon limit here for simplicity.  Keeping the 1 leads to a simple modification of the above, a geometry which may be interpreted as a non-extremal IIB Fluxbrane; it would be interesting to understand the relation between the NRCFT dual to our geometry and the worldvolume theory of the full stringy Fluxbrane.} of the usual D3 harmonic function and $f=1+g=1-{ \r^{4}_{H} \over \r^{4}}$ is the emblackening factor, together with the usual five-form fieldstrength supporting the $S^{5}$ and providing the 5d $cc$, 
$$
F^{(5)} = (1+*) \Omega_{5} ~\! d\theta\wedge d\phi \wedge d\psi \wedge d\mu\wedge d\chi
$$

Melvinizing this solution along the Hopf fibre is a 
straightforward application of the 
recipe described above, so let's jump to the chase and simply write down the final result (for completeness, the full computation is presented in an appendix).  In String Frame\footnote{The transformation to 10d Einstein Frame multiplies 
the metric by $e^{-{1\over 2}\Phi}=K^{1\over 4}$.}, using the coordinates $\{t,\xi,r\}$ introduced above, the result is
\be
\label{SchrBH}
ds^2 ={1 \over r^2 \KK} \(
- 2 f {\beta^2 \over r^2} dt^2 + 2 d\xi dt + {1-f\over 2} ( dt - d\xi)^2 \)
 + {1\over r^2} d\vec x^2  + {dr^2 \over r^2 f}  + ds_{\IP^2}^2 + {1 \over \KK} 
 \(d\chi + \CA\)^2.
\ee
where 
$(d\chi + \CA)$ and $ds_{\IP^{2}}$ are as above and 
$$
f = 1 - {r ^4 \over r_H^4} , ~~~~ \KK = 1 + {\beta^2 r^2 \over r_H^4} .
$$
In contrast to the $T=0$ solution and to the AdS black hole, the dilaton now has a non-trivial profile,
$$ 
\Phi = - \half \ln \KK, 
$$
while the Neveu-Schwarz two form takes a slightly different form,
$$ 
B = {\sqrt {2} \beta \over r^2 \KK} \( (1+f)  dt +(1-f) d\xi \) \wedge \( d\chi + \CA \)  
$$ 
The five form field strength is again unmodified by the Melvinizing.  

It is a simple but tedious exercise to verify that this is a solution to the full 10D IIB supergravity equations of motion.
Explicitly, it solves
$$ 
G_{\mu\nu} = \sum_{p=1,3,5} T^{(p)}_{\mu\nu} e^{ - \delta_{p}^{3} \Phi} ,
$$
where $T^{(p)}$ is the stress tensor for a minimally-coupled p-form field strength $H$,
$$
T_{\mu\nu}^{(H)}  = 
- {2 \over p (p+1) }  
\(
{1\over 4}  g_{\mu\nu} H^2
- {p \over 2} H_{\mu \cdots } H_{\nu}^{~\cdots}
\)
$$
and $T^{(1)}$ is the dilaton stress tensor.

There are many things worth noting about this solution.  
We focus first on the region near the horizon.
The component of the gauge field along the 
null killing vector normal to the horizon
(\ie\ $B_{\tau \phi} = A_\tau$)
vanishes at the horizon.
This is necessary to 
have a smooth euclidean continuation. 
The geometry contains a nice Rindler horizon 
with normal (and tangent) vector $\del_\tau$,
just as in the pre-Melvin hole  -- in particular, the 
would-be null-singularity living near $r\to\infty$ arising from compactification
of $\xi$ is lost behind the Schwarzschild horizon at $r=r_{H}$.
Near the horizon at $r= r_H$, 
it is useful to change coordinates to 
$$r =: r_H- R^2 ;$$
up to an irrelevant constant scale factor the metric looks like
\def\bbb{\kappa}
$$ 
ds^2 =  dR^2 - R^2 \bbb^2 d\tau^2 + {1\over r_H^2} d\vec x^2  + ...
$$
with 
$ \bbb \equiv {2 \over r_H}. $
In order for this to produce a smooth euclidean cigar geometry,
the euclidean time coordinate $\tau$
must be identified according to $ i \bbb \tau \simeq i \bbb \tau + 2 \pi $. 
This is the same result as for the pre-Melvin hole. 
Before Melvinizing, $\tau$ was also the asymptotic time coordinate,
and this gave a Hawking temperature $ T_H^{{\rm pre-Melvin}} = {\bbb \over 2\pi} = {1 \over \pi r_H} $.  
With the Schr\"odinger asymptotics, however, 
$ t = {1\over \sqrt 2} \( \tau + y\) $ is the natural time coordinate.
Different Hamiltonians imply different temperatures.
The euclidean continuation of our time coordinate $t$ must be identified according to
 $ i \bbb \sqrt{2} t \simeq i \bbb \sqrt{2} t + 2 \pi $,
and hence the Hawking temperature of our black hole is 
\be
\label{finallythetemperature}
T_H = {\sqrt 2 \over \pi r_H}.
\ee
%
Note that the horizon is unmodified by the Melvin procedure \cite{Gimon:2003xk};
only the asymptotics (and the relationship to the asymptotic coordinates and horizon coordinates) is changed.   
%

Next note the relative factor of $K$ between the $\IP^2$ part of the metric and the $\dz$ term: turning on a temperature has squashed the $S^5$ along the Hopf fibre (though, since $\K$ varies between $1$ and $1+\b^2$ between boundary and horizon, the squashing is gentle).  When we compactify to 5 dimensions, then, we should expect a non-trivial profile for the scalar field associated with this squashing 
mode.  Note, too, that this squashing breaks the isometry group of the sphere from $SU(4)$ to a subgroup,
and thus breaks the supersymmetry of the background accordingly\footnote{
For work on the supersymmetric generalization of the Schr\"odinger group
see \cite{Sakaguchi:2008rx}.}.  Entertainingly, T-dualizing on this fibre still leaves us with a squashed sphere, but the dual dilaton is now constant.  This gives a perhaps-simpler IIA description which may be convenient for various purposes.

This solution also admits a number of illuminating limits.  As before, taking $\b\to0$ effectively un-does the Melvinization, returning us to the ordinary black D3-brane solution with an identification.  Taking $T\to0$ sends this solution to the zero temperature solution found above, ie to 
$Sch^{2}_{5}\times S^{5}$. 

All of which raises the obvious question, {\em what is $\b$?}  At $T=0$, we saw that $\b$ could be scaled out of any physical question 
by 
a rescaling of coordinates, which 
can be thought of as a choice of units in the boundary theory. 
However, 
this is not the case at finite temperature.
$\b$ 
thus represents a physical parameter of the finite-temperature black hole embedded in asymptotically Schr\"odinger spacetime.  To anticipate what property of the spacetime/NRCFT this parameter represents\footnote{We will verify this through explicit computations of boundary thermodynamic quantities in the next section.}, look back at the Melvinization procedure.  In step (3), $\a$ turns on a mixing of the $y$ direction (which will eventually become the $\xi$ direction after boosting to the IMF) and angular momentum along the $S^{5}$.  In finite temperature AdS/CFT, angular momentum along the sphere translates into finite chemical potential for the conserved R-charge dual to the angular momentum current.  Combining the above, we should expect the CFT dual to this emblackened $Sch^{z=2}_{5}$ to have a finite $\xi$-momentum density, {\it aka} particle number density, which scales as some power of $(\beta / r_{H}^{m})$, where the factor of $r_{H}^{m}$ with $m>0$ is there to ensure that the density runs to zero as $T ={\sqrt{2}\over\pi r_{H}}\to0$ with $\b$ held fixed since, as we have seen, any finite $\beta$ is unphysical at zero temperature so cannot determine any physical quantity like density in the zero temperature limit.


\subsection{In Search of QCP: Finite Density at $T=0$}
\label{finitedensity}

The solutions found above have non-zero temperature and 
density.  However, sending $T\to0$ appears to send them {\em both} to zero.  This is bizarre, particularly in a non-relativistic theory in which particle-antiparticle annihilation is absent so that the number density should stay fixed as we take the temperature to zero.  We must be able to find a finite-density zero-temperature solution!  
And indeed it is useful to do so, 
since refrigeration techniques have reached the point 
that thermal effects on the cold atoms can be neglected for many purposes.

The answer was already implicit in the last paragraph of the previous subsection: 
the particle number density
of our Schr\"odinger black holes scales as some power of $\b /r_H^{m}$; the limit we should take, then, 
involves sending $r_H \to \infty$ (which removes the horizon and sends the temperature to zero) 
while taking $\b\to\infty$ to hold the density fixed.  
A little experimentation suggests that the proper limit is to scale $r_H^2 \sim \beta^{-1} \to \infty$, keeping $\Omega\equiv{\b \over r_H^{2}}$ fixed.  

%


\def\aa{\heartsuit}
To define the limit more precisely, we make the following replacements:
\be\label{scalings}
 t = \sqrt 2 \tilde \beta \aa \tilde t, ~~ \xi = { \tilde \xi  \over \aa \sqrt 2 \tilde \beta} , 
 ~~ r_H = \aa^{1/2} \tilde r_H, ~~ \beta = \aa \tilde \beta.\ee
We will show in Section 4 that the particle number density in these units is given by 
$\tilde{\rho} \propto \Omega^{2}\equiv{\tilde{\b}^{2} \over \tilde{r}_H^{4}}$, while
the rescaled temperature is $ \tilde T = { \sqrt 2 \over \pi \aa^{3/2} \tilde r_H}$.
To send $\tilde{T}\to0$ keeping finite $\tilde{\rho}$, we should take $\aa \to \infty$ holding objects with tildes fixed (we will drop the decorations at the end).
Note that (\ref{scalings}) includes the transformation
which allowed us, at zero temperature, to scale away any finite $\b$ (notably, in the present limit, $\b\to\infty$).

Ignoring the sphere directions for simplicity, the metric 
in the scaling limit $\aa \to \infty$ takes a pleasingly simple form,
\def\K{\kappa}
\be
\label{Tzeromunonzero}
ds^{2}_{\Omega\neq0} ={1\over r^{2}\K} \( -\frac{dt^2}{r^2} 
+2dtd\xi +{ \Omega^2 r^4}d\xi^{2}\) +  {d\vec{x}^{2}+dr^{2}\over r^{2}}
\ee
while the $B$ field takes the equally entertaining form,
$$
B={1 \over r^{2}\K}\(dt + 2 \Omega^2 r^{4}d\xi\)\wedge\dz
$$
and the dilaton remains non-trivial,
$$
\Phi=-\half\ln{\K},
$$
where $\K \equiv 1+\Omega^{2} r^{2}$.  The five-form, as usual, goes along for the ride.  Perhaps unsurprisingly given its pedigree, but surprisingly given its form (note that as $r\to\infty$, $r^{2}\K\sim r^{4}$, so the $\xi$ direction asymptotes to a finite radius controlled by $\Omega$), this background can be explicitly shown to solve the full equations of motion of IIB supergravity.  We will study the thermodynamics of this solution alongside that of the finite-temperature case in the next section.  

Note, that in our scaling limit,
the horizon has run off to $r = \infty$.  Happily, the null-singularity observed
before near $r\to\infty$ is absent thanks to the $r^{4}d\xi^{2}$ 
term -- finite density has cut off this singularity. 
However,
the dilaton still grows logarithmically.  
This means that the theory contains a
region of strong coupling 
in the IR part of the geometry,
somewhat similar to gravity duals 
of IR-strong gauge theories, such as 
Dp-branes with $p<3$ \cite{IMSY}.
It would be exciting to interpret
this scale-dependence of the string coupling
in terms of screening in the boundary theory
at finite density.

\subsection{An empty trap: Finite Temperature and Zero Chemical Potential}

If we can take the temperature to zero holding the chemical potential fixed, it stands to reason that we can take the chemical potential to zero holding the temperature fixed.  Indeed we can!  However, something rather stupid happens.  To see this, let's run the scaling of the last section backward, holding $T$ fixed but scaling $\Omega\to0$. It is easy to identify the proper scaling here: we must take $\b\to0$.  Sending the chemical potential to zero thus sends the solution back to the black D3-brane with which we began.  Of course, we have compactified the light-like $\xi$ direction, so what we really have is a DLCQ of AdS (and hence the breaking of the symmetry group from $SO(2,4)$ to its $Sch^{z=2}_{5}$ subgroup).  However, as an NRCFT, it is rather disappointing -- there is nothing in the trap.

As should by now be clear, the backgrounds we have been studying are intimately connected to DLCQs.  Fleshing out this connection will shed light both on the spacetime solutions themselves and on the NRCFTs to which they are dual.  The remainder of this section is thus devoted to an analysis of this connection.

\subsection{The Null Melvin Twist with finite $L_{\xi}$ as a modified DLCQ}

The Null Melvin twist has the great advantage of being a concrete tool with which to generate 
new IIB solutions from our tired old examples, and as such has been studied rather extensively.  However, at intermediate steps the solutions are far from simple, and the physical meaning of the procedure is rather opaque: what is the intrinsic relationship between the final and initial solutions?

\def\bDLCQ{{DLCQ$_{\b}$}}

Happily, there is another way of organizing the argument which is completely equivalent and which makes the connection between initial and final solutions manifest, 
following \cite{Bergman:2001rw}.  Let's start by studying the DLCQ of the original solution along the $\xi=(y - \tau)/\sqrt{2}$ direction by requiring all fields $\Phi$ to be invariant under translation along the light-cone $\xi$ direction, 
$$\Phi(\xi+L_{\xi})=e^{L_{\xi} J_{\xi}}\Phi(\xi)\stackrel{!}{=}\Phi(\xi),$$  
where $J_{\xi}=\p_{\xi}$ is the momentum generator on the light-like direction.  This produces the solution above at $\b=0$.  So: how do we introduce $\b$?  The answer is suggested by step (3) of the Melvinization, in which we re-diagonalized our symmetry generators to mix the spatial translation current $dy$ into the angular rotation current $d\phi \to d\phi+\a dy$; this replaces momentum along the (boosted) $y$-direction, $J_{y}$, with the sum of $J_{y}-\a J_{\phi}$, where $\a J_{\phi}=\p_{{1\over\a}\phi}$.   Boosting back to our original frame and taking the boost large turns this into $J_{\xi}-\half\a e^{\g}J_{\phi} \to J_{\xi}-\b J_{\phi}=J_{\b}$.  
To get the final solution with $\b\neq0$, then, we should perform a modified DLCQ of the original solution in which we orbifold not by a finite translation by $L_{\xi}J_{\xi}$, but by the modified current, $L_{\xi}J_{\b}$, \ie\ we should shift the light-cone momentum of every field by $\b$ times its charge under the $d\phi$ isometry.  We'll refer to this as a \bDLCQ.

The physical meaning of the generated solution is thus relatively straightforward: our Null Melvin Twist, {\it aka} the \bDLCQ, is a restriction 
of the original solution to modes with fixed light-cone- and 
angular- momenta\footnote{Note that for nonzero $\beta$, $J_{\b}$ is not actually light-like in the bulk.  So this is in general a DLCQ 
only from the point of view of the boundary field theory.}.
Note that the \bDLCQ\ of zero-temperature $\CN=4$ SYM has  a particularly trivial limit in which we hold $L_{\xi}$ fixed and scale $\b\to0$ -- this is just the usual DLCQ of $AdS_{5}$.  The fact that this system realizes the Schr\"odinger group as isometries is a direct consequence of the compactification of $\xi$.   Unfortunately, this limit ensures that the particle density of the groundstate is zero, so, while this does describe a NRCFT 
it describes the system at zero density only.
It is thus extremely important to preserve the parameter $\b$ if we want to describe something like ``fermions at unitarity'' rather than  ``{\em no} fermions at unitarity.''

\subsection{The \bDLCQ\ of $\CN=4$ SYM}

The virtue of the \bDLCQ\ prescription for our purposes is that it translates relatively easily 
into the dual field theory.
To wit, the NRCFTs dual to our 
Schr\"odinger spacetimes arise as \bDLCQ's of the boundary theories of the original solution.  Importantly, the current by which we orbifold is $J_{\b}=J_{\xi}-\b J_{R}$, where $J_{R}$ is the U(1) R-current dual to the isometry current on the $S^{5}$, $J_{\phi}$, we used in Melvinization.  For example, the NRCFT dual to Sch$^{z=2}_{5}$ is simply the \bDLCQ\ of the $\CN=4$ SU(N) SYM living on the boundary of the AdS$_{5}$ with which we started, with $J_{R}$ the trace of the cartan of the $SO(6)$ R-symmetry (corresponding to the Hopf fibration we used in Melvinization).

This result may also be derived via direct application of the Null Melvin Twist to the field theory as follows.
Start with the $\CN=4$ SYM theory and compactify it on a circle 
$y \simeq y + 2\pi L_y$ with all fields $\Phi$ required to satisfy the boundary conditions,
$$e^{L_{y} (\p_{y}-i\a q_{R})}\Phi(y) =  \Phi(y),$$  
where $q_{R}$ is the charge of $\Phi$ under the specified $U(1)$ subgroup
of the R-symmetry group.  This is equivalent to inserting an R-symmetry-valued
wilson line around the compact spatial direction \cite{Bergman:2001rw}.
For our special case, we chose the R-symmetry such that all three complex scalars in the ${\bf 6}$ of $SO(6)$ carry the same charge;
this corresponds to a shift on the Hopf circle.
Now boost the $y$ direction by $\g\to\infty$ to make it lightlike while scaling $\alpha \to 0$ such that
$ \beta = \half \a e^{\g} $ remains fixed.  
At weak 't Hooft coupling, this has the following effect.
On the potential terms in the Lagrangian it does nothing
because the R-symmetry is a symmetry.  
On $y$-derivative terms it amounts to the replacement
$\del_\xi\Phi \to \del_{\xi} \Phi -i \beta q_R \Phi $.
In terms of $\xi$-momentum, the net effect is to shift the moding of each field by a constant piece proportional to $\b$, ie
$$ L_{\xi} (i\ell-i\b q_{R}) = 2\pi i ~~~~ \Rightarrow ~~~~ \ell={2\pi N \over L_{\xi}} +\b q_{R}, $$
where $N$ is an integer.  This is precisely the \bDLCQ\ described above.

This theory seems remarkably simple, even moreso than the usual un-modified DLCQ of $\N$=4.  To understand why, recall that the usual DLCQ tells us to expand every field in the theory in modes along the light-like $\xi$ circle.  This leaves a KK tower of massive modes, plus a single level of massless modes -- the zero modes of $\p_{\xi}$ -- which must be treated with, if not respect, at least care.  The resulting theory is thus deliciously close to being non-relativistic, but the persistence of these zero modes reminds everyone that the theory is really Lorentz invariant.
To get a truly non-relativistic theory, we would like to lift these zeromodes.  
But since the upshot of the \bDLCQ\ is to
shift the moding of all fields by $\b$ times their R-charge, that is precisely what the \bDLCQ\ does.  More precisely, the only zero-modes surviving the \bDLCQ\ are those of R-scalars, ie of the vector bosons, and these must be dealt with carefully; among other things, they generate instantaneous interactions between the remaining non-zero modes of the matter fields.
The result is a theory with only nonrelativistic excitations, with the spectrum gapped by two mass scales, the KK scale $1/L_{\xi}$ and the new scale $\b$.  
In particular, something dramatic happens when
$\b \times ({ L_{\xi}\over 2\pi}) \in  \IZ$: $\Phi$ picks back up a zero mode (this is just the fact that the wilson line along the DLCQ circle has phase $\b L_{\xi}$).   $\b$ is thus playing the role of an IR regulator for the DLCQ zero modes generated by a wilson line.   Another curious feature of this scaling is that in order to excite a single KK mode, we need $\O({2\pi\over L_{\xi}\b})$ would-be-zero-modes; this suggests that there is an interesting regime where
$1 \ll \vev{N} \ll  {2\pi\over L_{\xi}\b}$ where we can drop the KK modes but we still have a well-regulated 
theory.  Understanding the interplay between these two scales in more detail, and especially to see it arise in the dual geometry, seems worthwhile; we leave such questions to future work.


\section{Black Hole Thermodynamics in Schr\"odinger Spacetimes}\scz

\subsection{Entropy}


%


The Bekenstein-Hawking entropy density of the black hole (\ref{SchrBH}) is
\be
\label{entropydensity}
 s \equiv {S \over L_1 L_2} = { 1\over 4 G_{10}}  L_y {\pi^3 \over r_H^3} R_A^8  = {1\over 4 G_5} L_y {  R_A^3 \over r_H^3 } .
 \ee
 Note that the dependence on $\beta$ cancels.
 To write this in terms of more physical variables, we need to relate $L_y$ to $L_\xi$.
%
What we mean by $L_y$ in the formula (\ref{entropydensity}) is 
the extent of the horizon in the $y$ direction when $\xi$ has period $ L_\xi$.  
To figure out what this is, one need only look
at the metric near the horizon, and plug in.
Near the horizon, the metric takes the form:
$$ ds^2 = dy^2 + ... = \half ( dt - d\xi) ^2 + ...$$
where $...$ is terms which vanish when we
ask about the invariant distance between two events 
separated only in the $\xi$ direction by an amount
$d\xi = L_\xi$.
Therefore:
$$ L_y = {1\over \sqrt {2} } L_\xi .$$

%
Using the standard parameter map of AdS/CFT
(which commutes with Melvinization)
$$ {R_A^8  \over 4 G_{10} }= {N^2 \over 2 \pi^4} ,$$
and the temperature (\ref{finallythetemperature}),
we have 
$$ s = {1\over 8} N^2 \pi^2 L_\xi T^3 . $$
For later comparison, it will be useful to note that 
in units where $ 16 \pi G_5 = 1$, we have 
\be
\label{entropyin5d}
 s =  L_\xi \pi^4 T^3. 
 \ee

\subsection{A comment on the correspondence for the stress tensor}

We begin with
a comment about the mysterious-seeming equation (34) of \cite{Son:2008ye},
which built on ideas of \cite{Son:2005rv}\footnote{We are grateful 
to Dominik Nickel and Pavel Kovtun
for help in appreciating this equation.}.
The last section of \cite{Son:2008ye}
 contains an assertion about which modes of the 
metric couple to which thermodynamic variables of the boundary theory,
which is supported by matching to a weakly coupled Lagrangian.
This expression can be understood more directly as follows.
In the standard AdS/CFT examples, fluctuations of the metric 
which have nice equations of motion are the ones with one upper and one lower index.
These are also  the components which couple directly to the boundary stress tensor
\cite{Liu:1998bu, Policastro:2002se}, \ie\
$$ I_{bdy} = \CO\(h^0\)+ \int_{\del M} \sqrt \gamma~ 2 \(T_{{\rm bdy}}\)^\mu_\nu h^\nu_\mu + \CO\(h^2\)$$
where $\gamma$ is the metric on the boundary.
Given $h^\mu_\nu$, to determine the perturbation of the metric
$ g_{\mu\nu} = g^{(0)}_{\mu\nu} + h_{\mu\nu}$, 
the right thing to do is to symmetrize:
$$ h_{\mu\nu} =\half \( g_{\mu\rho} h^{\rho}_\nu + g_{\nu \rho} h^{\rho}_\mu\). $$
Adding such a fluctuation to the zero-temperature ``schr\"odinger metric", and setting
$$A_0 = h^\xi_t, ~~~A_i = h^\xi_i , ~~~ \Phi = h^t_t, ~~~ B_i = h^t_i$$
gives Son's equation (34) to linear order in these fluctuations.
If $h^t_\xi$ is nonzero,
there is a nonzero fluctuation of $h_{\xi\xi}$.
For general $z$, and restoring factors of $\beta$, 
we find to linear order in the fluctuations
\be
\label{metricfluctuations}
 ds^2 = ds_0^2 + \( {A_0 \over r^2 } - 2 \beta^2 {\Phi\over r^{2z}} \) dt^2 
+ {\Phi(r) \over r^2 }d\xi dt  + 
\(   {A_i  dx^i \over r^2}  - { \beta^2 B_i dx^i \over r^{2z} } \) dt 
+  {B_i dx^i \over r^{2} } d\xi 
+h_{ij} dx^i dx^j + ... 
\ee

So $T^t_\xi$ is the number density of the field theory.
An analogy which is useful for understanding this point
is the following.  In IIB on $AdS_5 \times S^5$, 
considered as a ten-dimensional theory with a nine-dimensional boundary,
what is the meaning of $T_\mu^\phi$ and $T_\phi^\mu$,
components of the boundary stress tensor with indices in the sphere directions?
The answer is that they give R-current densities.
This is quite analogous to the statement that $T^t_\xi$
gives the number density, since in our correspondence the 
particle number density is the density of $\xi$-momentum,
just as the R-charge density is the density of momentum
around the $S^5$ directions.

Note that the interpretation of 
$\CT^\xi_t$ and $\CT^\xi_\xi$ 
remains mysterious.

\subsection{Expectation values of the stress tensor}

Consider any 
bulk theory where the matter lagrangian doesn't involve
derivatives of the metric.
If the boundary metric is flat,
the terms in the on-shell action which are linear in the metric fluctuations 
take the form \cite{Liu:1998bu}\footnote{In what follows
we studiously set the bulk coupling $\CK_5 = {1\over 16 \pi G_5}$ to one.}
$$ 
I_{bdy} = \CO(h^0) +\int_{\del M} h_\mu^\nu \( \Theta^\mu_\nu -  \Theta  \delta^\mu_\nu \) 
+ I_{bdy, ct} +\CO(h^2)$$
where
$ I_{bdy, ct} $ contains counterterms involving the matter fields.
$$\Theta_{\mu\nu} = \half\( D_\mu n_\nu + D_\nu n_\mu\) $$
is the extrinsic curvature of the boundary,
and $n^\mu$ is an inward-pointing unit normal vector to the boundary.
%
Taking $n^r = - \sqrt{g^{rr}}$,
the formula to extract the expectation value of the field theory stress tensor 
from the bulk data is
$$ \(T_{{\rm bdy}} \)^\mu_\nu = -2 \sqrt \gamma \( \Theta^\mu_\nu - \delta^\mu_\nu ( \Theta+a ) + ...\) $$
where $\gamma$ is the metric on the boundary (\ie\ the metric on a fixed-$r$ subspace),
and $a$ is a counterterm coefficient,
and .... is the contribution of other counterterms.
This quantity should have a finite limit as $r \to 0$ 
(\ie\ as it approaches the boundary).

In these nonrelativistic systems there is one further complication 
in the extraction of the expectation values of the field theory stress
tensor.
This is the fact that the description of nonrelativistic 
systems one finds here involves an extra dimension $\xi$,
whose momenta are associated with the conserved particle number.
Since the ordinary stress tensor of the nonrelativistic system,
which we will denote $\CT$,
is an operator of particle number zero (\ie\ it is 
of the form $\Psi^\dagger ... \Psi$),
it is related 
to the boundary stress tensor which depends on $\xi$ by extracting the zeromode.
This leads to an extra factor of $L_\xi$:
$$ \CT^\mu_\nu = -2 \lim_{r\to 0} L_\xi  \sqrt \gamma \( \Theta^\mu_\nu - \delta^\mu_\nu ( \Theta+a ) + ...\) $$

\bigskip

We evaluate the stress tensor expectation values
in terms of the five-dimensional description.
It would be a useful check to redo this calculation
in ten dimensions.
The boundary counterterms we include are 
$$ I_{ct} = \int_{bdy} d^{d+2}X ~\sqrt{\gamma} \(a_1 e^{ \alpha_1 \Phi} 
+ a_2 e^{ \alpha_2 \Phi} A^2 
+ a_3 e^{ \alpha_3 \Phi} A^4 
\), $$
where $A^2 \equiv A_{\alpha} A_{\beta} \gamma^{\alpha \beta}. $\footnote{
We thank Dominik Nickel for pointing out that the $A^4$ term can contribute.
}
Because of the asymptotic behavior of the solution,
we can replace the factors of $e^{ \alpha_i \Phi} $ with their 
Taylor expansion about the boundary $r=0$;
$\alpha_3$ will not matter.
Note that $\Phi$ here is a proxy for any 5d scalar quantity which 
behaves as $e^{- 2 \Phi } = K$.

We find that finite expectation values of the physical components of the stress tensor
require the addition of some extrinsic terms:
\be
\label{neumannizing}
I_{\rm ext} = \int_{\rm bdy} \sqrt \gamma n^r A^\mu F_{r\mu} a_4 e^{ \alpha_4 \Phi} .
\ee
Note that to the order at which this term contributes to the stress tensor and free energy,
we can rewrite
$$ a_4 e^{ \alpha_4 \Phi } = a_4 + a_4 \alpha_4 \Phi . $$
This term changes the boundary conditions on the massive gauge field
away
from purely Dirichlet \cite{Hawking:1995ap, Klebanov:1999tb}.
Using our result in appendix C that the coefficient of the $F^2$ term 
in the 5d lagrangian is $ e^{ - { 8 \over 3} \Phi} $,
we see that 
the special choice $a_4 = 1, \alpha_4 = - {8\over 3} $
implies {\it Neumann} boundary conditions on $A$.
Remarkably and mysteriously, it turns out that 
$ a_4 = 1$ 
is required for finiteness of $\CT^t_t$,
and 
$ \alpha_4 = - {8\over 3} $
is required 
for the first law of thermodynamics to be satisfied.
%


In a scale-invariant field theory with dynamical exponent $z$,
the energy density and pressure in thermal equilibrium are 
related by\footnote{For the special cases $z=1,2$ this is
shown in \cite{LL};
the formula for general $z$ was derived with
Pavel Kovtun.}
$$z \CE = d \CP .$$
Just like tracelessness of the stress-energy tensor of a 
relativistic CFT (the special case $z=1$), this relation arises as a Ward identity for conservation
of the dilatation current. 
For our case with $d=z=2$, this implies $\CE =\CP$.
We  constrain the counterterms to cancel
the divergences at $r\to 0$ and so that the Ward identity
is satisfied\footnote{
The conditions on the counterterms we find are: 
$$ a_1 = -6,  \alpha_1 = - {1\over 6} \( 2 a_2 - 4 a_4 - 2 \), a_4 = 1 $$
for finiteness, and the Ward identity requires
$$ 0 = 17 - 18 \alpha_1 - a_2 \( 6 + 10 \alpha_2 \) + 12 a_3 . $$
}. 
Identifying $\CE = - \CT^t_t$ and $\CP = \CT^i_i$, we find\footnote{In these expressions
we have divided out a common factor of $\CK = {N^2 \over 16 \pi^2} $ 
in all of the one-point functions.}
%
\be
\label{EandP}
 \CE = \CP  = {L_\xi \over 4} (\pi T)^4\(1+\aleph \delta^4\)  .
\ee
The numerical factor $\aleph$ depends
on counterterm coefficients which are not determined by 
finiteness of the energy, pressure, density, or by the Ward identity for scaling.
We will fix $\aleph$ below by demanding the first law of thermodynamics.
$\aleph$ will turn out to be zero, in agreement with \cite{Maldacena:2008wh,Herzog:2008wg}.
As a small check on our calculation, 
the action evaluated on the black hole solution satisfies
$$ T \(  I_{\rm bulk} + I_{\rm GH} +  I_{\rm ct} + I_{\rm ext} \)\big |_{{\rm on-shell}}  = \CP L_1 L_2 ,$$
as expected for the free energy in the grand canonical ensemble.  This equality is true 
of the regulated expressions for any choice of the counterterms.

Note that the thermodynamic potential densities $\CE, \CP$ in a system with dynamical
exponent $z$ should scale like $T^{d+z}$ times some function of 
the dimensionless ratio $\mu \over T$,
in agreement with our expressions (\ref{EandP}).
In our $z=2$ case,
the factor of $L_\xi$ makes up for the dimensions of the extra power of temperature.  


As discussed in the previous subsection,
the density is determined by $\vev{ \CT^{t}_\xi}$.  This gives
\be 
\label{densityresult}
\rho = 2 {L_\xi \over r_H^4 } = \half L_\xi (\pi T)^4. 
\ee

Note that the still-mysterious $T^\xi_\mu$ components of the stress 
tensor are still divergent.  That some components of the stress tensor
would remain divergent in holographic calculations with degenerate boundaries was anticipated in
\cite{Taylor:2000xf}.  The fact that the components 
which are hard to renormalize are precisely those whose
physical interpretation is unclear is heartening.

\subsection{Comments on chemical potential}
\label{subsecmu}

Son \cite{Son:2008ye} showed
that the mode $A_0$ of the metric 
in (\ref{metricfluctuations}) is the bulk field associated to
the boundary number density current.  
The expansion of the finite-temperature metric (\ref{SchrBH})
at the boundary gives
$$g_{tt} = - {2 \beta^2 \over r^4 } + {4 \beta^4 \over 3r_H^4 } {1\over r^2} + ... $$
Comparing with the parametrization of the fluctuations in (\ref{metricfluctuations}), we see that
$A_0 = { 4 \beta^4 \over 3 r_H^4} + \CO(r^2)$
in our background.
This suggests that $ { 4 \beta^4 \over 3 r_H^4}  = {4 \over 3} \delta^4$
determines the chemical potential for the number density in this background.
To extract more precisely the value of the chemical potential
indicated by these falloffs of $A_0$ 
requires a better understanding
of the couplings of these modes \cite{DominikPavel}.

\def\vv{\textswab{v}}

{\bf Note added in v2:} Following \cite{Maldacena:2008wh,Herzog:2008wg},
we can use a trick to determine the chemical potential
which makes precise the comments at the end of section 3.3.
The null killing vector at the horizon is $\vv \propto \del_\tau = {1\over \sqrt 2} \( \del_t - \del_\xi \)$.
If we normalize $\vv$ so that its component along the asymptotic time direction is unity,
$$ \vv = \del_t - \del_\xi ,$$
the temperature of the black hole is given by $T_H = {\kappa \over 2\pi}$;
the surface gravity $\kappa$ is defined as 
$$ \kappa^2 =- { 1\over 2}  \nabla_a \vv^b \nabla_c \vv^d g^{ab} g_{cd} .$$
This corroborates our earlier result that $ T_H = {\sqrt 2 \over \pi r_H}$.
Now, the fact that the null killing vector at the horizon
does not point only in the time direction
says that the ensemble to which the black hole 
contributes has a density matrix $\hat \rho= e^{ - {1\over T} \( \hat H - \mu \hat N \) }$;
this is the translation operator by which the euclidean geometry is identified.
In our $t, \xi$ coordinates, this gives 
\be 
\label{chemicalpot}
\mu = - 1. 
\ee

\subsection{Comments on the first law of thermodynamics}

The first law of thermodynamics should read
$$ \CE + \CP = Ts + \mu \rho .$$
Given the entropy density, thermodynamic relations 
determine $\mu \rho$ in a system with these symmetries.
From the Bekenstein-Hawking formula, we have an entropy density of the form
$$ s = c_1  L_\xi T^3 ,$$
where $c_1$ is a constant.
But the thermodynamic relation
$ s = {\del \CP \over \del T }$
implies
$$ \CP = {1\over 4} c_1 L_\xi T^4 + p_0(\mu) = {1\over 4 } Ts + p_0(\mu) $$
where the second term is temperature-independent but otherwise
thus-far undetermined.
The scale-invariance Ward identity, $ z \CE = d \CP $, then implies
$$\CE + \CP = \( {d \over z} + 1\) \CP $$
so
$$ \mu \rho = \CE + \CP - Ts = \( {1\over 4}  \( {d \over z} + 1\) - 1 \)  Ts +  {1\over 4} p_0 $$
For our case $d=z=2$, this gives
$$ \mu \rho = {1\over 2} c_1 L_\xi T^4 + {1\over 4 } p_0 $$

Using the thermodynamic potentials $ \CE, \CP, \rho$ extracted
from $\CT$, 
the enthalpy (the left hand side of the first law $ \CE + \CP = Ts + \mu \rho $)
is 
$$ 
\CE + \CP = \half L_\xi ( \pi T)^4 ( 1 + \aleph \delta^4 ) . 
$$
Using (\ref{densityresult}), (\ref{chemicalpot}) and (\ref{entropyin5d}), the right hand side is 
$$ Ts + \mu\rho = L_\xi ( \pi T)^4 - \half L_\xi ( \pi T)^4  .$$
Consistency of the first law therefore requires $\aleph = 0$,
and determines the integration constant $p_0(\mu) = 0 $.

%


\subsection{Thermodynamics in physical variables}

By rescaling $t \to t' = at, \xi \to \xi' = b \xi$,
we can change the chemical potential to a value with respect
to which it is possible to differentiate.  
In these new coordinates, we have 
$$ \mu' = - {b \over a}, ~~T_H' ={ \sqrt 2 b \over \pi r_H}, ~~\CE' = \CP' = {L_\xi \over r_H^4 } {1 \over ab} , ~~
\rho' = 2 {L_\xi \over r_H^4} {1\over b^2}, ~~s' = {L_\xi \over \sqrt 2 r_H^3 } {1\over b}.$$
The first law still checks.
In order to preserve the dispersion relation $ 2l \omega + \vec k^2 = 0$
(\ie\ to preserve the $g_{t\xi}$ metric coefficient),
we should set $ a = {1\over b}$.  
In retrospect, the dispersion relation with positive mass should be $ 2l \omega = \vec k^2 $;
this can be accomplished by setting instead $ a = - { 1\over b}$;
this will also make the energy density positive.

Making the substitution $a = -{1\over b}, b = \sqrt \mu,  {1\over r_H} = {\pi T \over \sqrt{ 2 \mu} }$, then, we have
$$ \CE = \CP = {1\over 4} {L_\xi ( \pi T)^4 \over \mu^2 }  , ~~~~~
\rho = \half {L_\xi (\pi T)^4 \over \mu^3}  .$$

A small check on this result is the following.
The free energy of a scale-invariant theory at finite temperature and chemical potential
can be written as 
$$ F = - V T^\alpha f\( \mu \over T\) ~.$$
The power $\alpha$ is determined by dimensional analysis,
and for general $z$ turns out to be $ \alpha = { d+ z \over z}$.
Note that this value implies that $ z \CE = d \CP$,
in agreement with the scale-invariance Ward identity.
%
%
Free nonrelativistic gases, both classical and quantum with either statistics,
in the grand canonical ensemble 
give $ \alpha = (d + 2 )/2 $ \cite{LL}.  
%
For $z=1$, $p = T^{d+1}f\({\mu\over T}\) $ is the familiar scaling
(\eg\ when $\mu \to 0$).  
The 
behavior of $\alpha$ at more general $z$
can be argued as follows.
With scaling exponent $z$, temperature (which is an energy),
scales with $z$ powers of inverse-length.  
Therefore $T^{1/z}$ scales with one power of inverse-length.  
The free energy density should scale with $ d+z$ powers of inverse-length
to make up for the scaling of $\int dt d^dx $.
This gives $ \alpha  = { d+ z \over z}$,
which agrees with the two familiar cases.

\section{Viscosity}\scz

In this section we will study the shear viscosity $\eta$
of the fluid described holographically by the metric (\ref{SchrBH}).
We will do this using the Kubo formula  
\be
\label{kubo}
 \eta = - \lim_{\omega \to 0} {1\over \omega} \Im G^R(\omega, \vec k=0), 
 \ee
where $G_R$ is the retarded two point function of the 
scalar mode of the stress tensor:
$$
G^R(\omega, \vec k=0)
= - i \int d^dx dt e^{ i \omega t} \theta(t)  \vev{ [ \CT_{xy}(t,\vec x), \CT_{xy}(0,0)] } .$$
Here we emphasize that 
the stress tensor is an operator with particle-number zero\footnote{We thank Pavel Kovtun for an extremely useful conversation
on this point.}:
\be
\label{zeromode}
\CT_{\mu\nu}(t, \vec x) \equiv \int_0^{L_\xi} d\xi ~T_{\mu\nu} (t, \vec x, \xi) 
\ee

It was argued in
\cite{Kovtun:2003wp} that very generally 
the linearized Einstein equation for 
$\phi \equiv h^x_y(u)e^{ -i \omega t}$ is the scalar wave equation in the same background.
The argument uses only the $SO(2)$ symmetry of rotations in the $xy$-plane;
this symmetry is preserved in our solution. 
We have also explicitly checked this statement using the ten-dimensional IIB 
supergravity equations of motion.

Note that unlike the familiar case of three spatial dimensions,
in our $d=2$ example there is no third dimension
in which to give momentum to $h^x_y$.
However, this momentum must be set to zero before taking the
$\omega \to 0$ limit in the Kubo formula,
and nothing is lost
for the purposes of studying the viscosity.

We will show in the remainder of this section that the familiar relation 
$$ {\eta \over s} = {1\over 4\pi} $$
also holds in these models.  
Note that the form of our metric violates the
hypotheses of the general theorem \cite{buchelliu}.
It would be interesting to see how much further the assumptions made there
can be relaxed.

\subsection{Scalar wave equation in the finite-temperature solution}

For convenience, 
we will discuss this problem in six-dimensional Einstein-frame
(\ie\ dimensionally reduce on the constant-volume $\IP^2$).
We show that the answer is frame-independent in the appendix.


\def\EE{{\large\textswab{e}}}
\def\ky{\textswab{q}_y}
The wave equation is
$$ \Box \phi = - g^{\mu\nu} k_\mu k_\nu \phi
+ {1\over \sqrt g} \del_u \( \sqrt g g^{uu} \del_u  \phi\) .$$
In this metric, 
$$ \sqrt g = {\sqrt K \over 2 u^3}. $$
We will study Fourier modes of the form:
$$ \phi(\tau, y, \vec x, u) = e^{ i {2\over r_H}\( -\EE \tau +  \ky y 
\) } f_K(u)  ~~,$$
\ie\ we have already set to zero the momentum in the spatial directions
and the squashed-sphere directions.
Note that $\EE, \ky$ are dimensionless variables,
measured in units of the temperature (times ${2\pi\over \sqrt 2} $),
\ie\ they are the gothic variables of \cite{Son:2002sd, 
Policastro:2002se, Policastro:2002tn}.
It will be crucial to distinguish $\EE$ from the variable
$\omega$ conjugate to the asymptotic time coordinate $t$.

The wave equation becomes
$$ 0 =  u^3  \del_u \( 
{ 4 f
\over
 u } \del_u f_K \)
- 
 \(
 - {u \over f
 } \EE^2 
+ \delta^2 (\EE-\ky)^2 + u \ky^2 \) f_K .$$
The indicial equation near the horizon 
arises from setting $f_K = (1-u)^\alpha$ 
and demanding that the most singular terms at $u=1$ cancel.
This gives
$$0 =  \alpha^2 + {\EE^2 \over 4}. $$
The solution obeying incoming-wave boundary conditions
at the horizon takes the form
$$ f_K(u)  = (1-u)^{ - i \EE/2} F_K(u) $$
where $F_K$ is analytic at $u=1$.
Next, to study the hydrodynamic limit, we can expand $F_K$ in a small-frequency expansion:
$$ 
f_K(u) =
(1-u)^{- i \EE/2}\( 1 + \EE F_1(u) + \ky F_2(u) +...
\);$$
here the ellipses denote terms of order $\EE^2, \EE \ky, \ky^2$ \footnote{Actually, the correct expansion treats $ \EE $ as the same order 
as $\ky^2$; this will not affect the viscosity calculation.}.
Plugging back into the wave equation, we find, just as
in the AdS black hole \cite{Policastro:2001yc,Son:2002sd,Policastro:2002se},
$$ F_1(u) = i \ln { 1 +  u \over 2 } , ~~~F_2(u) = 0.$$

Using $ g^{uu} = {4 u^2 f \over \sqrt{\KK}}$ and $\sqrt{-g} = {\sqrt{\KK} \over 
2 u^3 r_H^4}$,
this produces a flux factor
$$
-{\cal F} = {\cal K} \sqrt{-g} g^{uu} f_{-K}(u) \del_u f_K(u) 
= {\cal K } {2 (1 - u^2)\over u r_H^4} \( 
{ i \EE \over 4}{1 \over 1 - u} - {1\over 4} { i \EE 
\over 1+u}  \) + \CO\( \EE^2, \ky\EE, \ky^2 \) $$
where 
$$ {\cal K } =  {N^2 \over 16 \pi^2 } $$
is the normalization of the bulk action,
written here in terms of field theory variables.
It will cancel in $\eta/s$.  
We need the relationship between the momenta
associated to the horizon coordinates and asymptotic coordinates:
$$ 
\ky = {1\over \sqrt 2} \( \omega + l\) r_H, ~~
\EE = {1\over \sqrt 2} \( l- \omega\)r_H .$$
Note that we have restored factors of $ {1\over \pi r_H}$ 
in the definition of the $t, \xi $ momenta relative to the gothic momenta.
At the boundary $u=\epsilon$, the flux factor is therefore
$$
-{\cal F}|_{u=\epsilon} =  
{{ \cal K  }\over r_H^3 } 
\( 
{i \over 2} {
\omega - l\over \sqrt 2 }  
+ \CO\(\omega, l, k^2, \ky^2\)\).
$$ 
We dropped contact terms in this expression.
The real-time AdS/CFT prescription of \cite{Son:2002sd}
says that 
the retarded Green's function 
is obtained from the flux factor by
$$ 
G^R(\omega, \vec k =0) = 
- 2 {\cal F}|_{u=\epsilon}.$$

At this point, we pause to consider whose Green's functions we are studying.
The momentum-space correlator in the Kubo formula (\ref{kubo})
has had a factor of the volume of spacetime divided out by
translation invariance:
$$ 
G(\omega, \vec k) = {1\over VT} \int d^{d+1}x_1\int d^{d+1}x_2~ 
e^{ i k_1 \cdot x_1 + i k_2 \cdot x_2 } G(x_1, x_2) ;$$
the factor $  VT = L_1 L_2 T$ is $\delta^{d+1}(0)$ in momentum space.
As emphasized in equation (\ref{zeromode}), the field theory stress tensor 
is the zeromode in the $\xi$ direction of the operator
to which $h^{\mu}_\nu$ couples.
Therefore, when we relate
the two-point function of $T(t, \vec x, \xi)$ to 
the momentum-space Green's function $G^R$, 
we should not divide out by the associated factor of $L_\xi$:
$$  \vev{ [ \CT_{xy}(\omega, \vec k), \CT_{xy}(- \omega, -\vec k)] }
 = 
 \int d\xi_1 d\xi_2  \int d^d \vec x dt e^{ i \omega t - i \vec k \cdot \vec x}
   \vev{ [ T_{xy}(t , \vec x, \xi_1), T_{xy}(0,0, \xi_2)] }.$$

Putting this together,
the Kubo formula for
the viscosity then gives
$$ \eta = - \lim_{\omega\to0} {1\over \omega} \Im G^R(\omega, \vec k =0)
=  2 {\cal K}  L_\xi { 1 \over \sqrt 2 r_H^3} 
= {\pi L_\xi T^3 N^2 \over 32}. $$
Note that in $d$ spatial dimensions $\eta$ indeed has mass dimension $d$.
This is identical to the familiar $\CN=4$ answer
except for
{\it a)} the interpretation as the viscosity of a theory in two spatial dimensions,
and 
{\it b)} the factors of $1/\sqrt 2$ which
come from the relation between 
the asymptotic time coordinate and the coordinate which
becomes null at the horizon.

Taking the ratio $ \eta \over s$ reproduces the KSS value 
$$ {\eta \over s} = {1\over 4\pi}. $$

%

%

\section{Discussion}\scz

Our black hole lives in a space with very different
asymptotics from AdS.
There structure of the horizon, however,
is the same as that of the AdS black hole; this is guaranteed by the manner
in which it was constructed \cite{Gimon:2003xk}.
The calculation of the viscosity
is not obviously determined only by the geometry near the horizon.
However, the factors conspire mysteriously to preserve the viscosity ratio.
Our result, then, 
is some further indication that the membrane 
paradigm should be taken seriously.

In this paper we have focussed on an example with
dynamical exponent $z=2$ in $d=2$ dimensions,
which is related to the $\CN=4$ theory by a twisted version of discrete light cone quantization.
Work on constructing string theory realizations for 
critical phenomena with other values of $z,d$ is in progress.   

It would be interesting to find the black hole solution which asymptotes to
the NR metric with spherical spatial section, \ie\ the analog of the 
black hole in global coordinates in AdS.
The melvinization can't work quite the same if the starting point is
AdS in global coordinates, because the analog of $y$ is then 
an angular variable.

Having identified a zero-temperature background
with nonzero density, and its likely weak-coupling description,
we can calculate the Bertsch parameter
(see \eg\ \cite{zwerger}) for this theory.  The Bertsch parameter
is the cold-atoms analog 
of the famous $3 \over 4$-ratio of strong and weak coupling 
free energies in the $\CN=4$ theory.

The boundary field theory we are studying clearly
contains bosonic excitations, which carry charge under the 
number-density operator.  
There should be a chemical potential to temperature ratio
above which they simply Bose condense.  
In this regard, it would 
be interesting to Melvinize the Sakai-Sugimoto model \cite{Sakai};
it has a better chance of describing a system conaining
only fermionic atoms.

A nice check on our result for the viscosity
and our understanding of the thermodynamics 
of the solution will be the location of the 
diffusion pole in the shear channel of the stress-tensor
correlators \cite{connor}.

\vskip0.2in
\noindent
{\bf Note added:}
When this work was substantially complete, we learned that 
two other groups \cite{Maldacena:2008wh,Herzog:2008wg} had found results
which overlap with ours.

\vspace{1cm}
{\bf Acknowledgements}
We thank Jan de Boer, Daniel Grumiller, Hong Liu, 
Connor McEntee, Dominik Nickel,
Krishna Rajagopal, Martin Zwierlein
and especially Pavel Kovtun
for discussions and encouragement.
A.A. thanks the organizers and participants of the 2008 Banff Workshop on String Theory where this work was first presented, and the Aspen Center for Physics and the Denver International Airport for hospitality during the writing of this paper.
J.M. is grateful to the 2008 Amsterdam Summer Workshop on String Theory for 
hospitality.
The internet router at the Eden Hotel Amsterdam is responsible for all errors
in the paper.
This work was supported in part by funds provided by the U.S. Department of Energy
(D.O.E.) under cooperative research agreement DE-FG0205ER41360.


\begin{appendix}

\section{Details of Melvinization}

In this appendix, we review the Null Melvin Twist, as 
formalized in a seven-step dance in \cite{Gimon:2003xk}.

\subsection{Buscher Rules and Conventions}\scz 

\bea
g_{yy}' = {1\over g_{yy}}
~~~~~~&
g_{ay}' = {B_{ay}\over g_{yy}}
~~~~~~&
g_{ab}' = g_{ab} - {g_{ay}g_{yb}+B_{ay}B_{yb}\over g_{yy}}\\
\Phi' = \Phi -\half\ln{g_{yy}}
~~~~~~&
B_{ay}' = {g_{ay}\over g_{yy}}
~~~~~~&
B_{ab}' = B_{ab} - {g_{ay}B_{yb}+B_{ay}g_{yb}\over g_{yy}}
\eea

\subsection{The Hopf Vector on $\IP^{2}$}\scz

In constructing our solutions we were forced to pick an isometry direction along $S^{5}$.  A particularly convenient choice involved realizing $S^{5}$ as a Hopf fibration over $\IP^{2}$, which we now review to make your life easier than ours was (if you don't already know this stuff).  

The round metrics on $\IP^{n}$ and $S^{2n+1}$ may be elegantly expressed in terms of the left-invariant one-forms of SU(n).  For SU(3), these can be written in coordinates as,
$$ \s_{1} = \frac{1}{2} (\text{d$\theta $} \cos (\psi )+\text{d$\phi $} \sin (\theta ) \sin (\psi ))$$
$$ \s_{2} = \frac{1}{2} (\text{d$\theta $} \sin (\psi )-\text{d$\phi $} \cos (\psi ) \sin (\theta )) $$
$$ \s_{3} = \frac{1}{2} (\text{d$\psi $}+\text{d$\phi $} \cos (\theta )) $$
In terms of these 1-forms, the metrics on $\IP^{2}$ and $S^{5}$ may be written,
$$ ds^2_{\IP^{2}} = \text{d$\mu $}^2 +\sin ^2(\mu ) \left(\sigma_1^2+\sigma _2^2+\cos ^2(\mu ) \sigma _3^2\right) $$
$$ ds^2_{S^{5}} = \text{ds}_{\IP^2}^2 +\left(\text{d$\chi $} +\sin^2(\mu) \sigma _3 \right){}^2 $$
where $\chi$ is the local coordinate on the Hopf fibre and $\A = \sin^2(\mu)\sigma _3={\sin^2(\mu)\over 2}(\text{d$\psi $}+\text{d$\phi $} \cos (\theta ))$ is the 1-form potential for the kahler form on $\IP^2$ ($d\chi+\A$ is the vertical one-form along the Hopf fibration).  This explicit coordinate presentation is 
necessary to verify that our various solutions in fact solve the full 10d IIB supergravity equations of motion,
and to study the linearized equations of motion for the fluctuations.

\subsection{Constructing the finite temperature solution}\scz
We now walk through the melvinization of the black D3-brane in all its majesty.

{\bf Step 1}: 
We start with the black D3-brane solution,
$$
ds^{2} = {1\over h} \( -d\tau^{2}f +dy^{2}+d\vec{x}^{2} \) + h \({d\r^{2}\over f} + \r^{2}\[ds_{\IP^{2}}^{2} + (d\phi +\A)^2 \] \)
$$
where $h^{2}= 1+{R^{4}_{A} \over \r^{4}}$ is the usual D3 harmonic function and $f=1+g=1-{ \r^{4}_{H} \over \r^{4}}$ is the emblackening factor.  In what follows, nothing untoward will happen to the $d\vec{x}^{2}$, $d\r^{2}$ or $ds^{2}_{\IP^{2}}$ factors, so we'll drop those terms and reintroduce them after the dust settles.  The truncated metric is thus,
$$
ds^{2} = {1\over h} \( -d\tau^{2}f +dy^{2} \) + h \r^{2}(d\phi +\A)^2 
$$

\def\dz{{(d\chi +\A)}}
{\bf Step 2}:  Boost by $\gamma$, ie $\tau\to c\tau-sy$ with $c=\cosh(\gamma)$ and $c^{2}-s^{2}=1$:
$$
ds^{2} = {1\over h} \( -d\tau^{2}(1+gc^{2}) +dy^{2}(1-gs^{2}) +2d\tau dy(gcs) \) + h \r^{2}\dz^2 
$$

{\bf Step 3}: T-dualize along the $dy$ isometry using the Buscher rules listed above:
\bea
ds^{2} = -d\tau^{2}{f\over h(1-gs^{2})} +h\(   \r^{2}\dz^2   + dy^{2}{1\over 1-gs^{2}}  \)  \\
B=2dy\wedge d\tau\[{ -gcs \over 1-gs^{2} }\] ~~~~~~~~\Phi = \Phi_{0}-\half\ln\[{1-gs^{2}\over h}\]
\eea

{\bf Step 4}: Shift the local 1-form $d\chi$ to $d\chi + \alpha dy$ to give
$$
ds^2 = -d\tau^{2}{f\over h(1-gs^{2})} + h\(   \r^{2}\dz^2   + dy^{2}{1 +\r^{2}\a^{2}(1-gs^{2})\over 1-gs^{2}} +2dy\dz(\a\r^{2}) \)
$$
Note that $\a$ has dimensions of ${1\over {\rm length}}$.

{\bf Step 5}: T-dualizing back along $dy$ gives
\bea
ds^2 &=& -{d\tau^{2} \over h(1-gs^{2})}\[ f-{g^{2}c^{2}s^{2} \over 1 +\r^{2}\a^{2}(1-gs^{2}) } \] 
+{2dyd\tau \over h }\[{ gcs \over 1 +\r^{2}\a^{2}(1-gs^{2}) }\]      \non \\ 
&~&~~~+{dy^{2} \over h }\[{ 1-gs^{2} \over 1 +\r^{2}\a^{2}(1-gs^{2}) }\]   
+ h\r^{2}\dz^{2}\[{1\over 1 +\r^{2}\a^{2}(1-gs^{2}) }\]  \non\\
B&=&{\a\r^{2} \over 1 +\r^{2}\a^{2}(1-gs^{2})}\dz\wedge\[ gcs~\!d\tau +(1-gs^{2})dy\]   \non\\
\Phi &=& \Phi_{0}-\half\ln\[{1-gs^{2}\over h}\]  \non
\eea

{\bf Step 6\&7}: We now boost back by $-\gamma$ and take a double scaling limit $\alpha\to0$ with $\alpha c =\beta$ held fixed.  Since many terms do not survive this, it is easiest to do both steps at once and report only the result, adding back in all the terms we dropped in the first step,
\bea
ds^2 &=& {1\over h K}\[  -d\tau^{2}(1+\b^{2}\r^{2})f  +dy^{2}(1-\b^{2}\r^{2}f)  +2d\tau dy(\b^{2}\r^{2}f)  \]  \non \\
&~&~~~~ +{1\over h}d\vec{x}^{2}
+h \[{d\r^{2}\over f} + \r^{2}ds_{\IP^{2}}^{2} + {\r^{2}\over K}(d\chi +\A)^2 \]  \non \\
B&=&{2\b\r^{2} \over K} \dz\wedge( f~\!d\tau +dy)   \non \\
\Phi &=& \Phi_{0}-\half\ln{K} \non
\eea
Note that $\beta$ has dimensions of ${1\over {\rm length}}$.

{\bf Step 8}: 
Finally, we take the near-horizon limit, $h\to R^{2}_{A}/ \r^{2}$.  
To compare with the solutions of \cite{Son:2008ye, Balasubramanian:2008dm},
it is convenient to switch variables to the radial radial coordinate 
$${r\over R_{A}}={R_{A}\over\r}$$ 
in terms of which the boundary is at $r=0$ and the horizon at $r_{H}=R^{2}_{A}/R_{H}$.  
In terms of $r$ and the parameter $\Delta = \b R^{2}_{A}$ we have 
$$
\b^{2}\r^{2}={\Delta^{2}\over r^{2}}  ~~~~
h={r^{2}\over R^{2}_{A}}  ~~~~
f=1-{r^{4}\over r^{4}_{H}}  ~~~~
K=1+{\Delta^{2}r^{2}\over r^{4}_{H}},
$$
with the metric taking the form,
\bea
ds^2 &=& {R_{A}^{2}\over  r^{2}K}\[  -d\tau^{2}(1+{\Delta^{2}\over r^{2}})f  +dy^{2}(1-{\Delta^{2}\over r^{2}}f)  +2d\tau dy({\Delta^{2}\over r^{2}}f)  \right. \non \\
&~&~~~~~~~~ \left. +K d\vec{x}^{2}
+ K{dr^{2}\over f} + r^{2}\(K ds_{\IP^{2}}^{2} + (d\chi +\A)^2\) \]  \non \\
B&=&2\Delta{R^{2}_{A} \over r^{2}K} \dz\wedge( f~\!d\tau +dy)   \non \\
\Phi &=& \Phi_{0}-\half\ln K \non .
\eea
Between the boundary and the horizon, $K$ varies smoothly between $1$ and $1+{\Delta^{2}\over r^{2}_{H}}$.   Importantly, the surface $r=r_{H}$, where $f\to0$ and $B_{t}\to0$, remains a non-singular null horizon.  Near the horizon, $\p_{\tau}$ is a timelike killing vector which is perpendicular to the null geodesics which span the horizon.  We thus have a non-rotating black hole with 
$\Omega_{H}=0$.  This might seem somewhat miraculous, since the geometry is not static but, like Kerr, only stationary, and so we might reasonably expect a Killing horizon outside the black hole.  In fact, this construction, which preserved the near-horizon geometry at each step, had built into it that the horizon would be irrotational (and, in particular, have no additional killing horizon).  We could introduce rotation by starting with a bifurcate killing horizon surrounding an ergosphere -- \ie\ by starting with a rotating black D3 -- but, since we will exploit the unbroken rotational symmetry of our solution to compute the viscosity, we'll leave this generalization to later consideration. 

The upshot is that we have a two-parameter family of finite-temperature solutions labeled by the $r_{H}$ and $\Delta$ defined in units of $R_{A}$.  This family has two simple and familiar limits, $\Delta\to0$ and $r_{H}\to\infty$.  Taking $\Delta\to0$, which sends $K\to1$, is easily seen to return us to the non-extremal black D3-brane solution with which we began.  

Taking $r_{H}\to\infty$, by contrast, takes us to the globally non-singular Schr\"odinger geometry.  To see this directly, it is useful to work in light-cone coordinates $t=(y-\tau)/\sqrt{2}$ and $\xi=(y+\tau)/\sqrt{2}$, in terms of which the solution becomes,
\bea
ds^2 &=& {R_{A}^{2}\over  r^{2}K}\[  -{2\Delta^{2}\over r^{2}}fdt^{2}  + 2dtd\xi   -{g\over 2}(dt-d\xi)^{2}  
+K d\vec{x}^{2} + K{dr^{2}\over f} + r^{2}\(K ds_{\IP^{2}}^{2} +(d\chi +\A)^2\) \]  \non \\
B&=&\rt~\!\Delta{R^{2}_{A} \over r^{2}K} \dz\wedge( [1+f]~\!dt +[1-f] d\xi)   ~~~~~~~~
\Phi = \Phi_{0}-\half\ln K \non .
\eea
In the limit $r_{H}\to\infty$, which takes $f\to1$ and $K\to1$, the metric reduces to, 
\bea
ds^2 &=& {R_{A}^{2}\over  r^{2}}\[  -{2\Delta^{2}\over r^{2}}dt^{2} + 2dtd\xi + d\vec{x}^{2} + dr^{2}  \] + R^{2}_{A} ds_{S^{5}}^{2}   \non \\
B&=&2\rt~\!\Delta {R^{2}_{A} \over r^{2}} \dz\wedge dt    ~~~~~~~~
\Phi = \Phi_{0} \non ,
\eea
which, upon compactifying on the $S^{5}$, is the Schr\"odinger geometry with 
$z=2, d=2$.  Studying the finite-$r_{H}$ solution near $r\ll r_{H}$ gives the same result.  We have thus embedded a black hole in an asymptotically Schr\"odinger spacetime.

One final set of coordinates will  be useful in the computations below.  In terms of the dimensionless quantities $u=r^{2} / r^{2}_{H}=R^{2}_{H}/\r^{2}$, $\delta=\Delta/r_{H}=\b R_{H}$ and  $\mu=R_{A}/r_{H}=R_{H}/ R_{A}$, the solution takes the form,
\bea
ds^2 &=& {\mu^{2}\over u K}\[  -{2\delta^{2}\over u}f~\!dt^{2}  + 2dtd\xi   -{g\over 2}(dt-d\xi)^{2}  +K d\vec{x}^{2} \right. \non \\
&~&~~~~~~~~~~ \left. +{K R^{2}_{A}\over 4 \mu^{2} u f}du^{2} + {u R^{2}_{A}\over\mu^{2}} \( K ds_{\IP^{2}}^{2} + (d\chi +\A)^2\) \]  \non \\
B&=&\rt~\!\delta{\mu  R_{A} \over u K} \dz\wedge( (1+f)~\!dt +(1-f) d\xi)   \non \\
\Phi &=& \Phi_{0}-\half\ln K \non .
\eea
where
$$
f=1-u^{2}  ~~~~
K=1+\delta^{2} u,
$$
These variables simplify many of the computations.

\section{Frame (in)dependence of the viscosity calculation}

After compactifying to $D$ dimensions,
the string frame metric is related to the $D$-dimensional
Einstein-frame metric by 
the Weyl rescaling 
$$  g^{E,D}_{\mu\nu} =e^{ 4 \Phi \over D-2}  g^{({\rm str})}_{\mu\nu}  .$$
In our solution, the dilaton is 
$$ e^{ 2\Phi} = {1\over \KK} $$ 
so we have
$$  g^{E,D}_{\mu\nu} = \KK^{{2 \over 2-D}}  g^{({\rm str})}_{\mu\nu}  $$
In the special case $D=10$, this says
$  g^{E,10}_{\mu\nu} = \KK^{{1\over 4}}  g^{({\rm str})}_{\mu\nu}  .$

Now consider the wave equation in a conformal frame reached
by an arbitrary power of $\KK$, where the metric is:
$$ g^a_{\mu\nu} = \KK^a g^{({\rm str})}_{\mu\nu}. $$
We have
$$ \det g^a = K^{D a } \det g^{({\rm str})} , ~~~
\sqrt{g^a} = {K^{5a-1} \over 2 u^3 r_H^4 } {\rm vol_{10-D}}  $$
where ${\rm vol_{10-D}}$ is the constant volume
of the compact dimensions, which will scale out of the wave equation.

The wave equation for a scalar in this background is
$$ \square \phi = 
{1\over K^{aD/2 } } 2 u^3  K\del_u \( { K^{-a} 4 u^2 f } 
{ K^{aD/2-1} \over 2 u^3 } \del_u \phi \)+ ... $$
$$ = {4 u^3  \over K^{aD/2 - 1} } \del_u \(
{K^{a \({D \over 2} - 1\) - 1 } f \over u } \del_u \phi \) + ... $$
The einstein-frame condition above
says that in $D$-dimensional Einstein frame,
$ K^a = e^{ - {4 \Phi \over D-2}}  $ 
which says 
$$ K^{a \({D \over 2} - 1\) - 1 } = 1. $$
So we see that in einstein frame, in whatever number
of dimensions we want to live in, say 10 or 6,
the factor $\KK$ does not appear in the wave equation.

This in turn implies that the viscosity is independent of $\delta \equiv {\beta \over r_H}$.

\section{Comments on
reduction to five dimensions}

\def\rebus{\Gamma} 
\def\C{\nu}
Let $\Gamma = - \half  \ln \KK$; this is the profile for 
both the 10d dilaton and the KK scalar associated to the Hopf direction.
The following two equations are true:
$$ 0 = - \del_\mu \( \sqrt g F^{\mu\nu} e^{ (\C-3)\rebus }  \) 
+ z(z+d) \sqrt{g} e^{ ( 3 \C-1) \rebus} A^\nu $$
$$ 0 = 16 \del_\mu\( e^{ ( 3 \C-1) \rebus} \sqrt g g^{\mu\nu} \del_\nu \rebus \) 
+ 
\sqrt{g} \( e^{ (\C-3)\rebus } F^2 + 2 z (z+d) e^{ ( 3 \C-1) \rebus} A^2 \) $$
where 
$$ e^{2\rebus} \equiv {1\over \KK} $$
$$ A = {2 \beta \over r^2 \KK} \( f d\tau + dy \)  $$
and
$$ds^2 = 
\KK^{\C}  {1\over r^2 \KK} \( 
- \( 1 + {\beta^2 \over r^2 } \) f d\tau^2  
- {\beta^2 f \over r^2}  2 dy d\tau 
+ \( 1 - {\beta^2 \over r^2 } f \) dy^2 
+ \KK d\vec x^2 + \KK {d r^2 \over f r^2 } \)
$$ 
These are the respective equations of motion for $A_\nu$ and $\Phi$ for a
five-dimensional action
of the form
$$ S_5 = \int d^5 x \sqrt g \( R - c_1 e^{  a_1 \Phi + b_1 \sigma } \(\del\Phi\)^2 
-c_2 \( {1\over 4}  e^{ a_2 \Phi + b_2 \sigma} F^2 + {m_A^2\over 2} e^{a_3 \Phi + b_2 \sigma} A^2 \) \) + \dots$$
with $m_A^2 = {z (z-d) \over L^2} $ as usual, and 
$ a_1 + b_1 = a_3+b_3 = 3 \nu - 1, a_2 = \nu - 3$
and $ a_2 = a_3$.
$ \nu = {1\over 3} $ is 5d Einstein frame.
Here $\sigma$ is the other scalar arising from the KK reduction.
The $\dots$ indicate terms that do not depend on $\Phi, A$.
We have not yet been able to determine the rest of the action.

\end{appendix}


\end{document}